\documentclass[prd,aps,twocolumn,a4paper,floatfix,showpacs,nofootinbib,reprint]{revtex4-1}

\usepackage{graphicx}
\usepackage{amsmath,amssymb}
\usepackage{physics}
\usepackage{subfigure}
\usepackage{appendix}
\usepackage[table]{xcolor}
\usepackage{hyperref}
\usepackage{booktabs} 
\usepackage{longtable} 
\usepackage{enumerate}
\definecolor{mangotango}{rgb}{1.0, 0.51, 0.26}

\def \Hz	{\mathrm{Hz}}

\def \imrxphm {\mathrm{IMRPhenomXPHM}}
 
\def \flow	{f_{\mathrm{low}}}

\def \fhigh	{f_{\mathrm{high}}}

\def \lalinf {\textsc{LALInference}}

\begin{document}

\title{Testing general relativity using higher-order modes of gravitational waves from binary black holes}

\author{Anna Puecher$^{1,2}$}
\author{Chinmay Kalaghatgi$^{1,2,3}$}
\author{Soumen Roy$^{1,2}$}
\author{Yoshinta Setyawati$^{1,2}$}
\author{Ish Gupta$^{4}$}
\author{B.S.~Sathyaprakash$^{4,5,6}$}
\author{Chris Van Den Broeck$^{1,2}$}

\affiliation{${}^1$Nikhef -- National Institute for Subatomic Physics, 
Science Park 105, 1098 XG Amsterdam, The Netherlands}
\affiliation{${}^2$Institute for Gravitational and Subatomic Physics (GRASP), 
Utrecht University, Princetonplein 1, 3584 CC Utrecht, The Netherlands}
\affiliation{${}^3$Institute for High-Energy Physics, University of Amsterdam, Science Park 904, 
1098 XH Amsterdam, The Netherlands}
\affiliation{${}^4$Institute for Gravitation and the Cosmos, Department of Physics, 
Penn State University, University Park, PA 16802, USA}
\affiliation{${}^5$Department of Astronomy and Astrophysics,  
Penn State University, University Park, PA 16802, USA}
\affiliation{${}^6$School of Physics and Astronomy, Cardiff University, Cardiff, CF24 3AA, 
United Kingdom}

\date{\today}

\begin{abstract}
Recently, strong evidence was found for the presence of higher-order modes in the gravitational
wave signals GW190412 and GW190814, which originated from compact binary coalescences with 
significantly asymmetric component masses. This has opened up the possibility of new tests of 
general relativity by looking at the way in which the higher-order modes are related to the basic signal. 
Here we further develop a test which assesses whether the amplitudes of sub-dominant harmonics
are consistent with what is predicted by general relativity. To this end we incorporate a 
state-of-the-art waveform model with higher-order modes and precessing spins into a Bayesian parameter
estimation and model selection framework. The analysis methodology is tested extensively through simulations. 
We investigate to what extent deviations in the relative amplitudes of the harmonics will
be measurable depending on the properties of the source, and we map out correlations 
between our testing parameters and the inclination of the source with respect to the observer. 
Finally, we apply the test to GW190412 and GW190814, finding no evidence for violations of general 
relativity.
\end{abstract}

\maketitle

\section{Introduction}
The Advanced LIGO \cite{TheLIGOScientific:2014jea} and Advanced Virgo \cite{TheVirgo:2014hva} 
gravitational wave (GW) observatories have by now detected 90 candidate signals from 
coalescing binary black holes 
\cite{Abbott:2016blz,Abbott:2016nmj,TheLIGOScientific:2016pea,LIGOScientific:2018mvr,Abbott:2020niy,LIGOScientific:2021djp},
binary neutron stars \cite{TheLIGOScientific:2017qsa,Abbott:2020uma}, and neutron star-black hole 
systems \cite{LIGOScientific:2021qlt}. A battery of tests of general relativity (GR) 
were performed \cite{TheLIGOScientific:2016src,LIGOScientific:2018jsj,Abbott:2018lct,LIGOScientific:2019fpa,LIGOScientific:2020tif,LIGOScientific:2021sio},
including tests of the spacetime dynamics as inferred from the binary coalescence process 
\cite{Arun:2006yw,Arun:2006hn,Yunes:2009ke,Li:2011cg,Li:2011vx,Agathos:2013upa,Agathos:2013upa,Meidam:2017dgf}. 

Recently, strong evidence was obtained for the presence of higher-order modes \cite{Blanchet:2008je} 
in the gravitational wave signals GW190412 and GW190814 
\cite{LIGOScientific:2020stg,LIGOScientific:2020zkf,Roy:2019phx}, which were emitted by coalescing 
binary compact objects with significantly different component masses. Measuring these sub-dominant 
harmonics of the basic signal enables more precise measurements of the source parameters, and can allow 
for stronger constraints on certain deviations from GR \cite{VanDenBroeck:2006qu,VanDenBroeck:2006ar}. 
Several tests of GR that directly probe the harmonic structure for binary black hole (BBH)
coalescences\footnote{Given the low mass of the lighter component of GW190814 ($\simeq 2.6\,M_\odot$), 
there is a possibility that it was a signal from a neutron star-black hole rather than a 
binary black hole coalescence \cite{LIGOScientific:2020zkf}, but studies based on the known 
properties of neutron stars make a BBH origin much more likely \cite{Essick:2020ghc,Tews:2020ylw}. 
For the purposes of this paper we will assume that GW190814 came from a BBH coalescence.} were proposed in 
\cite{Dhanpal:2018ufk,Islam:2019dmk,Kastha:2018bcr,Kastha:2019brk}. These fall into two categories. In 
the first case, one tests the phase evolution, e.g.~by
testing for deviations in the way parameters like the chirp mass and symmetric mass ratio 
enter into the expressions for the different harmonics \cite{Dhanpal:2018ufk}. This kind of test has 
already been applied to GW190412 and GW190814 in Ref.\,\cite{Capano:2020dix}. A second test looks for 
anomalies in the amplitudes of the sub-dominant modes \cite{Islam:2019dmk}; the latter test is the 
focus of this paper.

Specifically, defining $h(t) \equiv h_+(t) -i h_\times(t)$ with $h_+$, $h_\times$ the two polarizations, 
the GW signal from a coalescing binary can be written as
\begin{equation}
\label{eq:HOM}
  h(t; \iota, \phi_0, \vec{\lambda} ) = \sum_{\ell=2}^{\infty}\sum_{m=-\ell}^{\ell} 
  Y_{-2}^{\ell m}(\iota, \phi_0) \, h_{\ell m}(t; \vec{\lambda}),
\end{equation}
where the $Y_{-2}^{\ell m}$ are spin-weighted spherical harmonics of weight $-2$, 
$(\iota, \phi_0)$ indicate the direction of the radiation in the source frame, and 
$\vec{\lambda}$ collects all other parameters in the problem. The latter are the total mass
$M \equiv m_1 + m_2$ (with $m_1$, $m_2$ the component masses), the mass ratio $q \equiv m_1/m_2$ 
(where we assume $m_1\geq m_2$), the dimensionless spin vectors $\mathbf{S}_1$ and $\mathbf{S}_2$
at some reference time $t_{\rm ref}$, a reference phase $\varphi_{\rm ref}$, and the luminosity 
distance $D_{\rm L}$. Taking
the contribution with $\ell = 2$, $m = \pm 2$ to constitute the fundamental mode in the signal, the 
test of GR considered here follows Ref.\,\cite{Islam:2019dmk} to allow for deviations in the amplitudes of the 
higher-order modes:  
\begin{eqnarray}
&& h(t; \iota, \phi_0, \vec{\lambda} )  \nonumber\\
&& =   \sum_{m = \pm 2} Y_{-2}^{2 m}(\iota, \phi_0) \, h_{2 m}(t; \vec{\lambda}) \nonumber\\
&& + \sum_{\rm HOM} \sum_{m=-\ell}^{\ell} 
(1 + c_{\ell m})\,Y_{-2}^{\ell m}(\iota, \phi_0) \, h_{\ell m}(t; \vec{\lambda}),
\end{eqnarray}
where HOM stands for the $\ell$ labels of the higher-order modes. 
The $c_{\ell m}$ are free parameters, to be measured together with all other parameters in the 
problem; the case where GR is valid corresponds to $c_{\ell m} = 0$. Since 
$h_{\ell\, -m} = (-1)^\ell h^\ast_{\ell m}$, we set $c_{\ell\, -m} = c_{\ell m}$. Here we will perform 
parameterized tests where the $c_{\ell |m|}$ are allowed to vary one by one, as in the
phase-based tests performed in \cite{TheLIGOScientific:2016src,LIGOScientific:2018jsj,Abbott:2018lct,LIGOScientific:2019fpa,LIGOScientific:2020tif}, 
and we will focus on modes that will usually be the strongest, namely 
the ones with $(\ell, |m|) = (3,3)$ and $(\ell, |m|) = (2,1)$. We will not only 
perform parameter estimation, as was done in Ref.~\cite{Islam:2019dmk}, but also
model selection; as we shall see, 
the latter will be of particular importance here.

The observed strengths of the higher harmonics are set by the total mass $M$, 
the inclination angle $\iota$, and the relative mass difference $\Delta \equiv (m_1 - m_2)/M$ 
\cite{Blanchet:2008je,Blanchet:2013haa}. One aim
of this paper is to investigate to what extent deviations in amplitudes of the harmonics 
can be determined depending on the values of these parameters, in terms of both parameter 
estimation and model selection. Secondly, 
when performing tests that allow for non-zero $c_{\ell m}$, there will be correlations 
between these and the angular parameters, notably $\iota$, which will affect both the measurability of 
the deviations from GR and the shapes of the posterior distributions. We will  
map out this interplay, which is necessary to interpret the results of our tests. Finally, 
for the first time we apply this test to GW190412 and GW190814.

The rest of this paper is structured as follows. In Sec.~\ref{sec:basics} we 
recall the basic properties of higher harmonics, together with the waveform model
we will use. In Sec.~\ref{sec:setup} we set up
the Bayesian analysis framework used in this study, and explain our choices for simulated signals (or injections), 
which will be used to understand the behavior of our analysis depending on the properties
of the GW source. Sec.~\ref{sec:results} shows the results of our simulations and of 
measurements on GW190412 and GW190814. A summary and conclusions are provided in 
Sec.~\ref{sec:conclusions}.

\section{Properties of higher harmonics and waveform model}
\label{sec:basics}

\begin{figure}[t]
    \centering
    \includegraphics[width=0.48\textwidth]{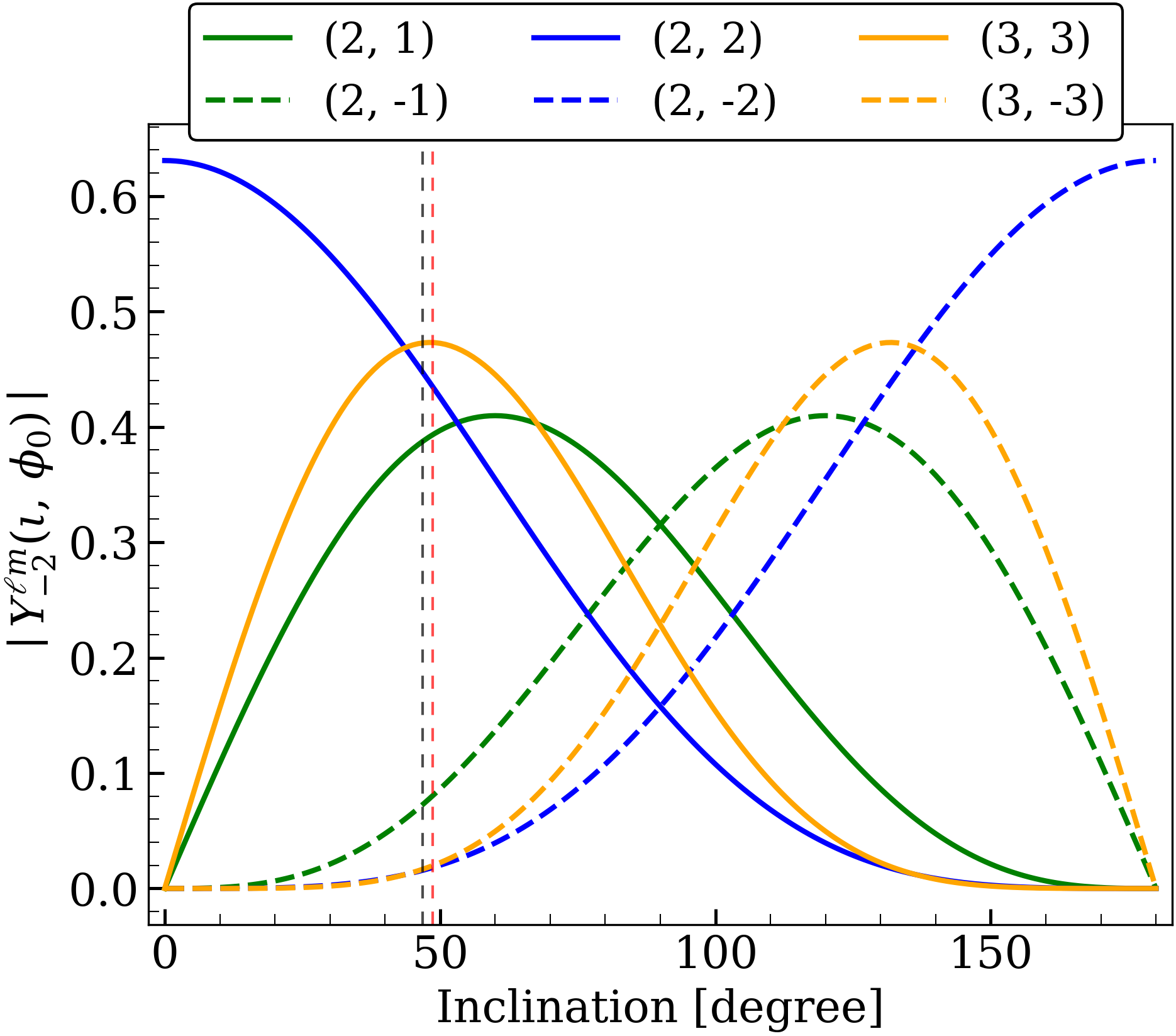}
    \caption{The absolute values of spin-weighted spherical harmonics of weight $-2$ as function of 
    the inclination angle $\iota$. The vertical lines indicate the peak likelihood values of 
    $\iota$ for GW190412 (black dashed) and GW190814 (red dashed), located at 
    $\simeq 47^\circ$ and $\simeq 49^\circ$, respectively 
    \cite{LIGOScientific:2020stg,LIGOScientific:2020zkf}.}
    \label{fig:Ylm_curve}
\end{figure}

Let us start by recalling some properties of the harmonics $h_{\ell m}$ in Eq.~(\ref{eq:HOM}), which we 
will need to interpret the results in subsequent sections. 
In doing so we limit ourselves to qualitative statements, 
mostly referring to the inspiral regime; for explicit dependences on the parameters in the problem we 
refer to Refs.\,\cite{Blanchet:2008je,Blanchet:2013haa}. The salient features relevant to us here are:
\begin{itemize}
\item At zeroth post-Newtonian order (0PN) in amplitude there is the harmonic with $\ell = |m| = 2,$ which is the most dominant of all multipole modes. 
\item At 0.5PN order in amplitude, harmonics with $(\ell, |m|) = (2,1), (3,3), (3,1)$ appear. In this 
paper we will be the most interested in the $(2,1)$ and $(3,3)$ harmonics, since the $(3,1)$ harmonic 
is suppressed with respect to the others due to its small overall numerical prefactor. For 
purposes of testing GR we will also not consider harmonics that only appear at higher PN order.
\item The $(2,1)$ and $(3,3)$ modes are proportional to the relative mass difference $\Delta = (m_1 - m_2)/M$, so that they are more prominent 
for systems with a higher value of $q = m_1/m_2$. 
\item The fact that the harmonics enter the polarizations through the spin-weighted spherical 
harmonics $Y_{-2}^{\ell m}(\iota, \phi_0)$ causes their prominence to depend sensitively 
on the inclination angle $\iota$, as illustrated in Fig.~\ref{fig:Ylm_curve}. For systems that 
are ``face-on'' ($\iota = 0$) or ``face-off'' ($\iota = 180^\circ$), only the dominant harmonic is 
visible. The sub-dominant harmonics on which we will focus on in this work are strongest 
around $\iota \simeq 50^\circ$ and $\iota \simeq 130^\circ$. In the figure we also indicate the 
peak likelihood values of $\iota$ for GW190412 and GW190814.
\item Finally, the observed power in the sub-dominant modes relative to that in the $(2,2)$ mode 
increases with the total mass, due to a combination of beyond-leading-order contributions to their amplitudes,
how much of the signal is in the detectors' sensitive band, and the shape of the noise 
power spectral density $S_n(f)$.
\end{itemize}

\begin{figure}[t]
    \includegraphics[width=0.48\textwidth]{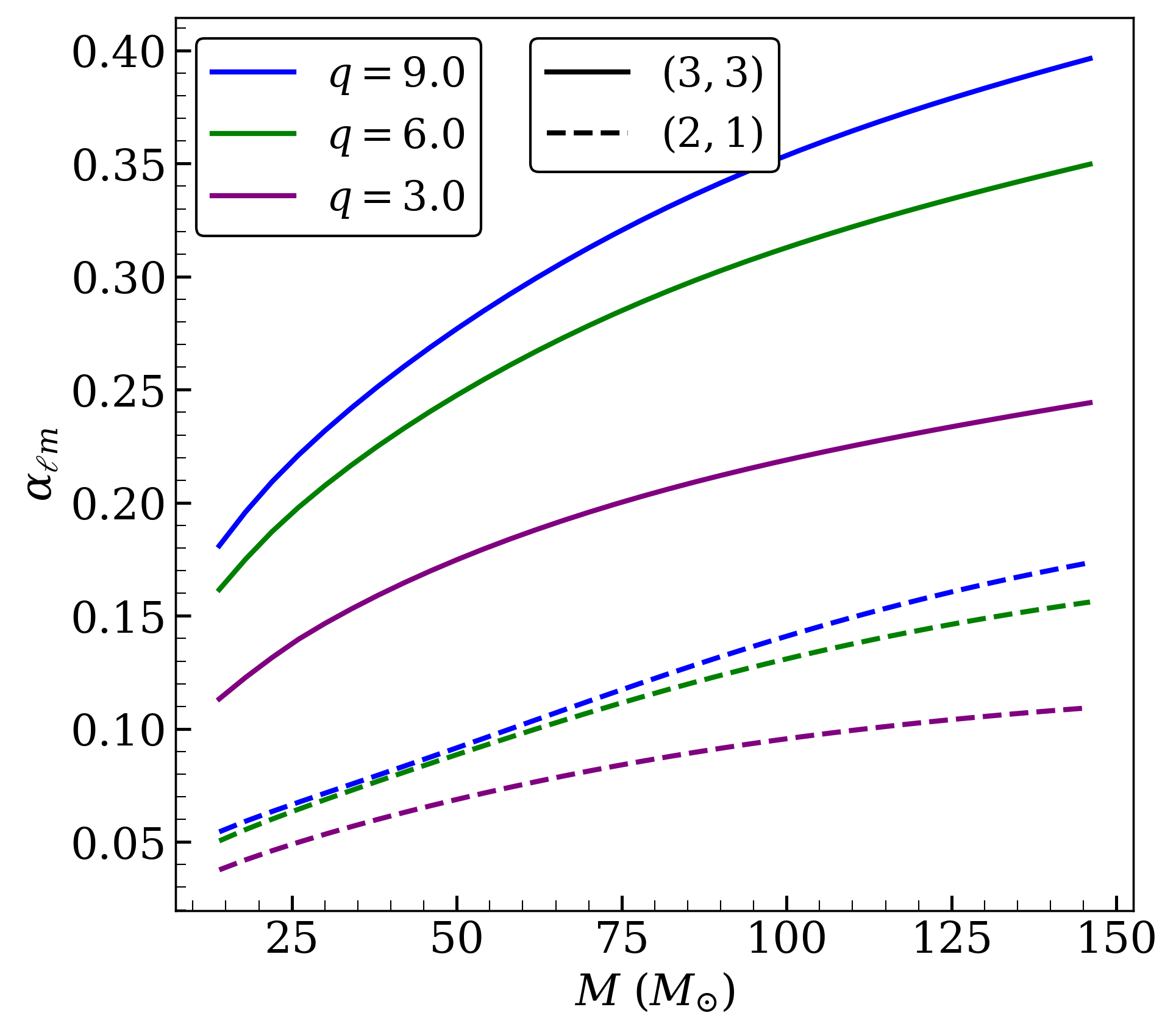}
    \caption{The relative signal power in the real part of $h_{\ell m}$ for 
    some of the higher-order modes with respect to the 
    dominant $(2, 2)$ mode, as a function of the total mass $M$ of the binary, for three different values
    of the mass ratio $q$ and assuming Advanced LIGO at design sensitivity.}
    \label{fig:various_modes_snr}
\end{figure}

To make the latter point more concrete, let us define 
the quantities 
\begin{equation}
\alpha_{\ell m} \equiv  \int_{\flow}^{\fhigh}  \frac{ \abs{ \tilde{ h}_{\ell m}(f; \vec{\lambda})  }^2 }{S_n(f)} 
\: df  \Big/ \int_{\flow}^{\fhigh}  \frac{ \abs{ \tilde{ h}_{22}(f; \vec{\lambda})  }^2 }{S_n(f)} \: df  ,
\end{equation}
where $\tilde{ h}_{\ell m}(f; \vec{\lambda})$ is the Fourier transform of the 
real part of $h_{\ell m}(t; \vec{\lambda})$, and $S_n( f )$ denotes the one-sided detector 
noise power spectral density, which we take to be the one for Advanced LIGO at design sensitivity 
\cite{TheLIGOScientific:2014jea}.  
The integrals are evaluated from a lower cut-off frequency $\flow=20\,\Hz$ to an upper cut-off frequency
$\fhigh = 2048\,\Hz$, which amply suffices for the kinds of signals considered in this paper. 
The waveform model is taken to be the most up-to-date phenomenological inspiral-merger-ringdown
model $\imrxphm$~\cite{Pratten:2020ceb, Ramos-Buades:2020noq}, which incorporates harmonics with
$(\ell, \abs{m}) = (2, 2), (2, 1), (3, 3), (3, 2), (4, 4)$ modes, as well as effects of 
spin-induced precession. Fig.~\ref{fig:various_modes_snr} shows the dependence of the 
$\alpha_{\ell m}$ on total mass $M$ and mass ratio $q$, for $(\ell, m) = (2,1), (3,3)$, where
for simplicity we have focused on binaries composed of non-spinning black holes.

\section{Analysis framework and setup of simulations}
\label{sec:setup}

We now explain our data analysis methodology for measuring source parameters and to rank  
hypotheses based on the available detector data. Next we will detail the choices made for simulations
that were performed to understand the response of the analysis framework to possible 
violations of GR in the amplitudes of different harmonics.

\subsection{Analysis framework}

Consider detector data $d$ and a hypothesis $\mathcal{H}$, where, for practical purposes, the
latter corresponds to a waveform model $\tilde{h}(f; \vec\theta)$; in our case this could be the GR model 
for binary black hole coalescence, or one that allows for deviations from GR in the amplitudes of one of the 
harmonics. Then, in a Bayesian setting, measuring the parameters $\vec\theta$ of the source amounts
to obtaining the \emph{posterior probability density} $p(\vec\theta | d, \mathcal{H})$. From
Bayes' theorem, 
\begin{equation}
p(\vec{\theta} | d, \mathcal{H}) = \frac{p( d | \vec{\theta}, \mathcal{H}) 
\: p(\vec{\theta} | \mathcal{H} )}{ p(d|\mathcal{H}) }, 
\label{Bayes}
\end{equation}
where the \emph{evidence} $p(d|\mathcal{H})$ for the hypothesis $\mathcal{H}$ 
is given by
\begin{equation}
p(d|\mathcal{H}) = \int d\vec{\theta}  
\: p( d | \vec{\theta}, \mathcal{H}) \: p(\vec{\theta} | \mathcal{H} ).
\label{Evidence}
\end{equation} 
In the above, $p(\vec\theta | \mathcal{H})$ is the \emph{prior probability density}, and 
the \emph{likelihood} $p( d | \vec{\theta}, \mathcal{H})$ takes the form
\cite{Veitch:2009hd}
\begin{equation}
p( d | \vec{\theta}, \mathcal{H})
\propto \exp\left[ -\frac{1}{2} \langle d - h(\vec\theta) |  d - h(\vec\theta) \rangle \right],
\label{Likelihood}
\end{equation}
where the noise-weighted inner product $\langle \, \cdot \, | \, \cdot \, \rangle$ 
is given by 
\begin{equation}
\langle a | b \rangle 
\equiv 4 \Re \int_{f_{\rm low}}^{f_{\rm high}} \frac{\tilde{a}^\ast(f)\,\tilde{b}(f)}{S_n(f)}df.
\label{Innerproduct}
\end{equation}

Eq.~(\ref{Bayes}) together with Eqs.~(\ref{Evidence})-(\ref{Innerproduct}) allow us 
to calculate the posterior probability density $p(\vec\theta | d, \mathcal{H})$ from the data.
The posterior probability density $p(\theta_k|d, \mathcal{H})$ for a particular parameter
in $\vec\theta$ is obtained by integrating out all the other parameters $\vec\xi$
in $\vec\theta = (\theta_k, \vec\xi)$:
\begin{equation}
p(\theta_k | d, \mathcal{H}) = \int d\vec\xi\,p(\theta_k, \vec\xi | d, \mathcal{H}).
\end{equation}

Additionally, we will want to rank hypotheses: the GR hypothesis $\mathcal{H}_{\rm GR}$ 
versus hypotheses $\mathcal{H}_{\rm NonGR},$ which allow one of the $c_{\ell m}$ to be non-zero.
To this end we calculate \emph{Bayes factors}, or ratios of evidences,  
\begin{equation}
\mathcal{B}^{\rm NonGR}_{\rm GR} \equiv \frac{p(d|\mathcal{H}_{\rm NonGR})}{p(d|\mathcal{H}_{\rm GR})},
\end{equation}
where $p(d|\mathcal{H}_{\rm NonGR})$ and $p(d|\mathcal{H}_{\rm GR})$ are obtained using 
Eq.~(\ref{Evidence}), taking $\mathcal{H}$ to be $\mathcal{H}_{\rm NonGR}$ or 
$\mathcal{H}_{\rm GR}$, respectively. In practice it is usually convenient to 
focus on the logarithm of the Bayes factor, $\ln \mathcal{B}^{\rm NonGR}_{\rm GR}$, 
as will also be done here.

It will also be important to consider the loudness of a signal as it appears in a detector. 
The optimal \emph{signal-to-noise ratio} (SNR) is defined as 
$\rho \equiv \langle h(\vec\theta) | h(\vec\theta) \rangle^{1/2}$. For a network of 
detectors, the combined optimal SNR is obtained by summing in quadrature the SNRs 
in the individual detectors.

Finally, for estimating the evidence integrals as in Eq.~(\ref{Evidence}), and obtaining
samples for posterior density distributions $p(\vec\theta | d, \mathcal{H})$, we
used nested sampling \cite{Skilling:2006gxv,Veitch:2009hd} as implemented in the
$\lalinf$ package \cite{Veitch:2014wba} of the LIGO Algorithms Library (LAL) 
software suite \cite{lalsuite}.

\subsection{Setup of the simulations}

To understand the response of our analysis pipeline to GR violations in mode amplitudes
with various strengths, we add simulated signals, or injections, to synthetic stationary, Gaussian noise
for a network of Advanced LIGO and Virgo detectors following the predicted noise spectral 
densities at design sensitivity \cite{TheLIGOScientific:2014jea,TheVirgo:2014hva}. Since 
higher-order modes are more prominent for larger total masses, we will start by considering heavier
BBH systems. Later in the paper we will analyze the real GW events GW190412 and GW190814 to look for GR violations.  To this end, for lower-mass systems we will perform injections whose GR parameter values and SNRs are set to the maximum-likelihood values obtained from analyses on these events 
that assumed GR to be correct. Specificallly:
\begin{itemize}
\item We will inject signals with $M = 65\,M_\odot$ and 
$M = 120\,M_\odot$, for mass ratios $q = 3, 6, 9$. Here the inclination angle is fixed to 
be $\iota = 45^\circ$, and the network SNR to 25.
\item For GW190412-like injections, $M = 46.6\,M_\odot$, $q = 4.2$, $\iota = 47^\circ$, and
the network SNR is 19.8. 
\item For GW190814-like injections, $M = 27.6\,M_\odot$, $q = 9.3$, $\iota = 49^\circ$, 
and the network SNR is 25. 
\end{itemize}

We also need to choose values for the deviation parameters $c_{33}$ and $c_{21}$ in the injections.
Since the $(3,3)$ mode will tend to be the strongest (see Fig.~\ref{fig:various_modes_snr}), 
we can expect smaller values of $c_{33}$ to lead to detectable GR violations than for 
$c_{21}$, where ``detectable'' can be taken to mean that the 90\% credible region of the posterior 
density function has support that excludes zero. We found that, at least for the higher masses listed above, 
the following choices constitute examples ranging from non-detectability to easy detectability 
of the GR violations:
\begin{itemize}
\item $c_{33} = 0.5, 1.5, 3$.
\item $c_{21} = 1, 3, 6$.
\end{itemize}
Hence these are the values for which we will show results in the next section.

\begin{figure*}
\centering
\includegraphics[width=0.47\linewidth]{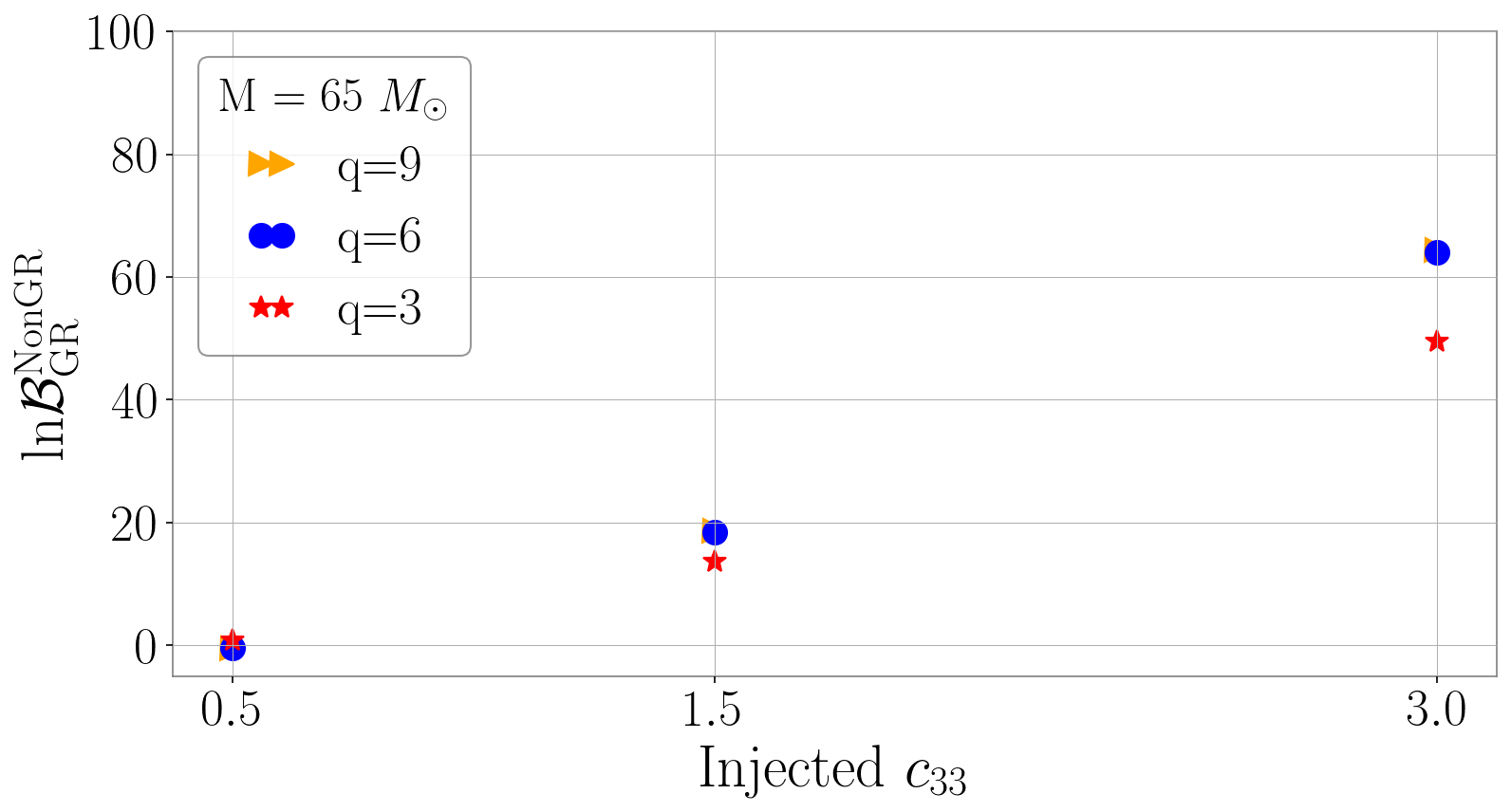}\hfill
\includegraphics[width=0.47\linewidth]{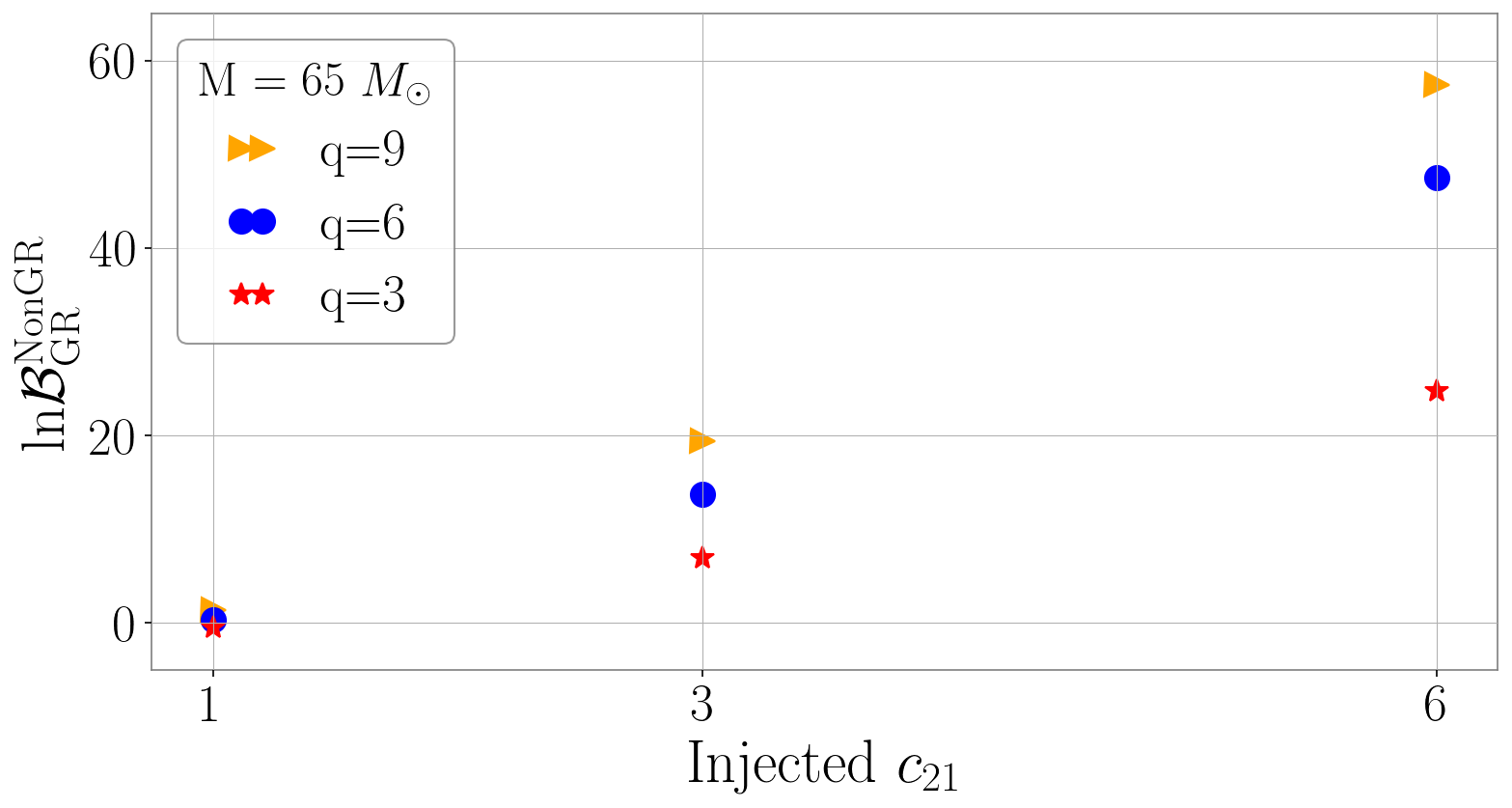}\hfill
 \includegraphics[width=0.47\linewidth]{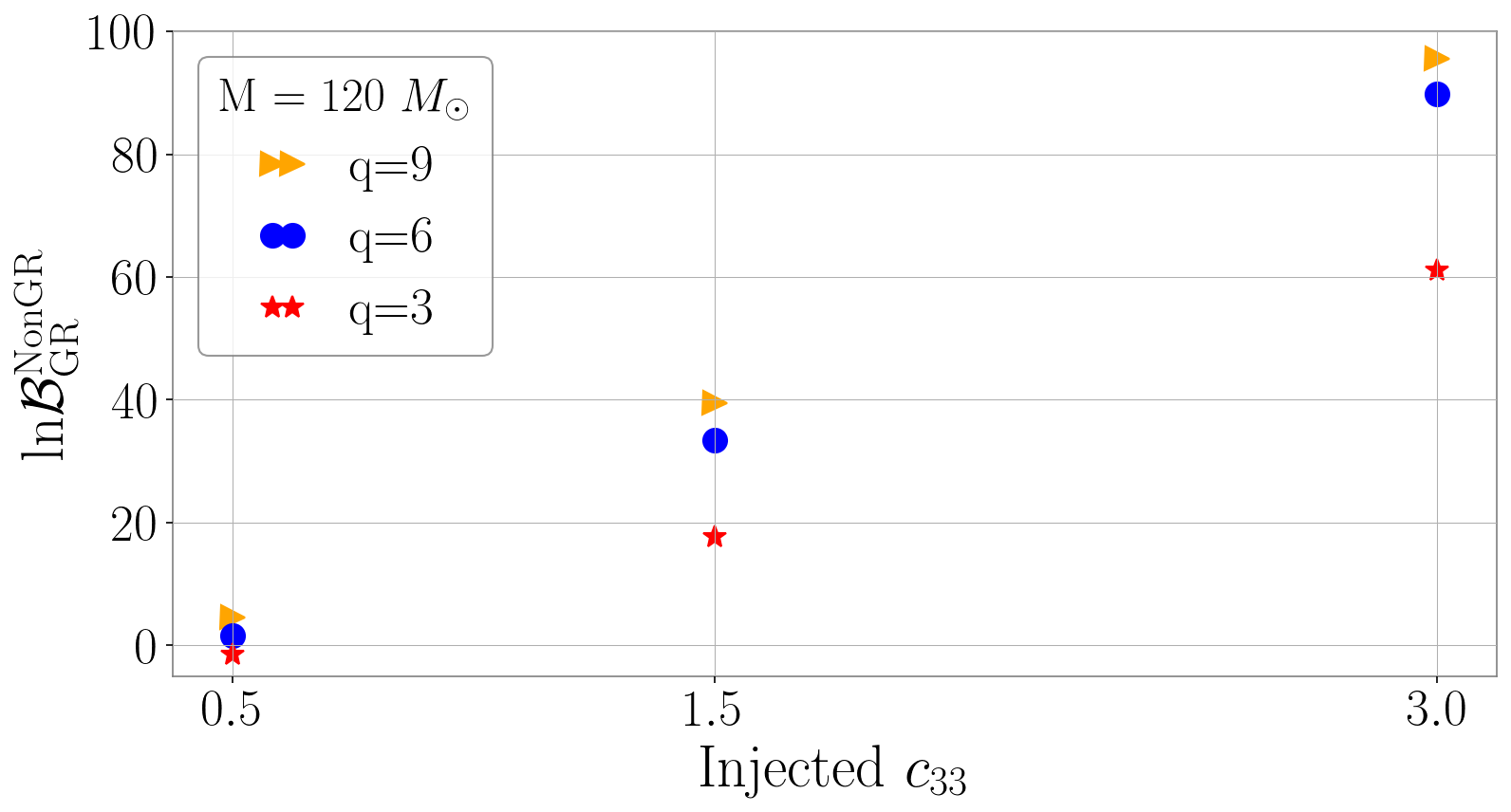}\hfill
  \includegraphics[width=0.47\linewidth]{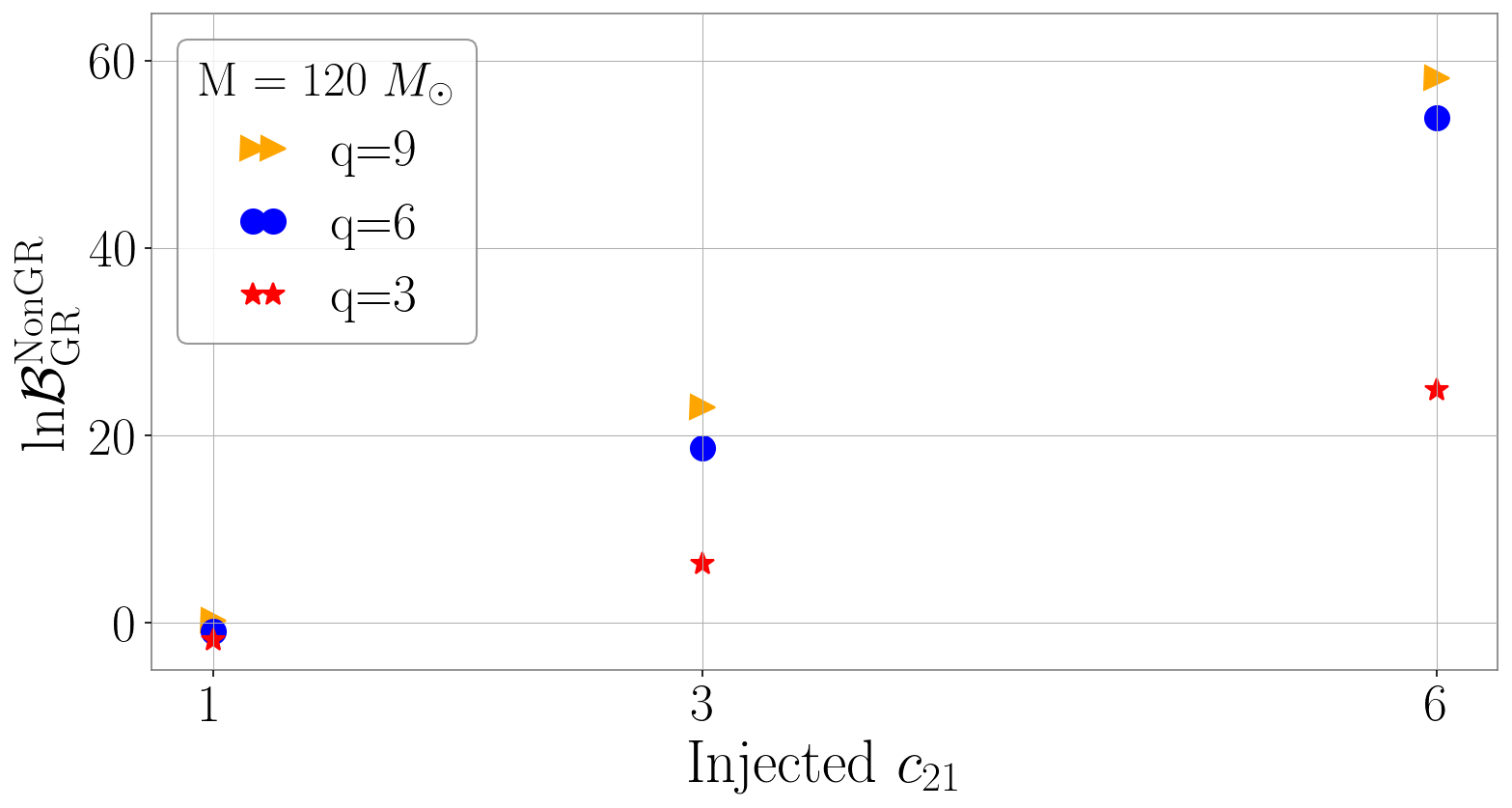}\hfill
 \caption{$\ln\mathcal{B}^{\rm NonGR}_{\rm GR}$ 
 for $M=65\,M_{\odot}$ (top row) and $M=120\,M_{\odot}$ (bottom row), for different mass ratios $q$ 
 indicated by the differently shaped markers. The horizontal axes 
 show the injected values of $c_{33}$ (left column) and $c_{21}$ (right column). In each case, 
 the non-GR hypothesis has the corresponding $c_{\ell m}$ as free parameter.}
 \label{fig:bayes_factor_difference}
\end{figure*}

\section{Results of simulations, and analyses of GW190412 and GW190814}
\label{sec:results}

We now describe the results for our simulations, as well as for the real events GW190412 and
GW190814, in terms of parameter estimation and hypothesis ranking. In doing so, it will be useful
to make a distinction between the more massive BBHs ($M = 65, 120\,M_\odot$), the 
injections with parameters similar to those of the real events, and of course the real events
themselves.

\subsection{More massive binary black holes}

Let us first look at results for injections with $M = 65\,M_\odot$ and $M = 120\,M_\odot$. To have an easier overview it 
is convenient to first look at the behavior of log Bayes factors, $\ln\mathcal{B}^{\rm NonGR}_{\rm GR}$, 
which we do in Fig.~\ref{fig:bayes_factor_difference}. The trends are as follows:
\begin{enumerate} 
\item As expected, for a larger injected $c_{\ell m}$, the log Bayes factor is larger. The cases 
$c_{33} = 0.5$ and $c_{21} = 1$ lead to $\ln\mathcal{B}^{\rm NonGR}_{\rm GR}$ that tend to be 
consistent with zero, meaning that the data are not sufficiently informative to clearly distinguish between
hypotheses. However, starting from $c_{33} = 1.5$ or $c_{21} = 3$, the $\ln\mathcal{B}^{\rm NonGR}_{\rm GR}$
are significantly away from zero, and as will be seen in terms of parameter estimation below, 
here the GR deviations tend to be detectable. 
\item Higher values of $M$ lead to higher $\ln\mathcal{B}^{\rm NonGR}_{\rm GR}$, consistent 
with there being more power in the higher-order modes relative to the $(2,2)$ mode; see 
Fig.~\ref{fig:various_modes_snr}.
\item Again as expected, on the whole a larger mass ratio $q$ tends to lead to a higher 
$\ln\mathcal{B}^{\rm NonGR}_{\rm GR}$, consistent with there being more power in the 
higher-order modes. We do see that the $\ln\mathcal{B}^{\rm NonGR}_{\rm GR}$ tend to differ 
less between $q = 6$ and $q = 9$ than between $q = 3$ and $q = 6$; in fact, for $M = 65\,M_\odot$
and $c_{33}$, the log Bayes factors for the higher two values of $q$ are nearly equal. Again pointing to  
Fig.~\ref{fig:various_modes_snr}, we note that the cases $q = 6$ and $q = 9$ are closer to 
each other than to $q = 3$ in terms of the power present in higher-order modes.
\end{enumerate}

\begin{figure*}
\centering
\includegraphics[width=0.98\linewidth]{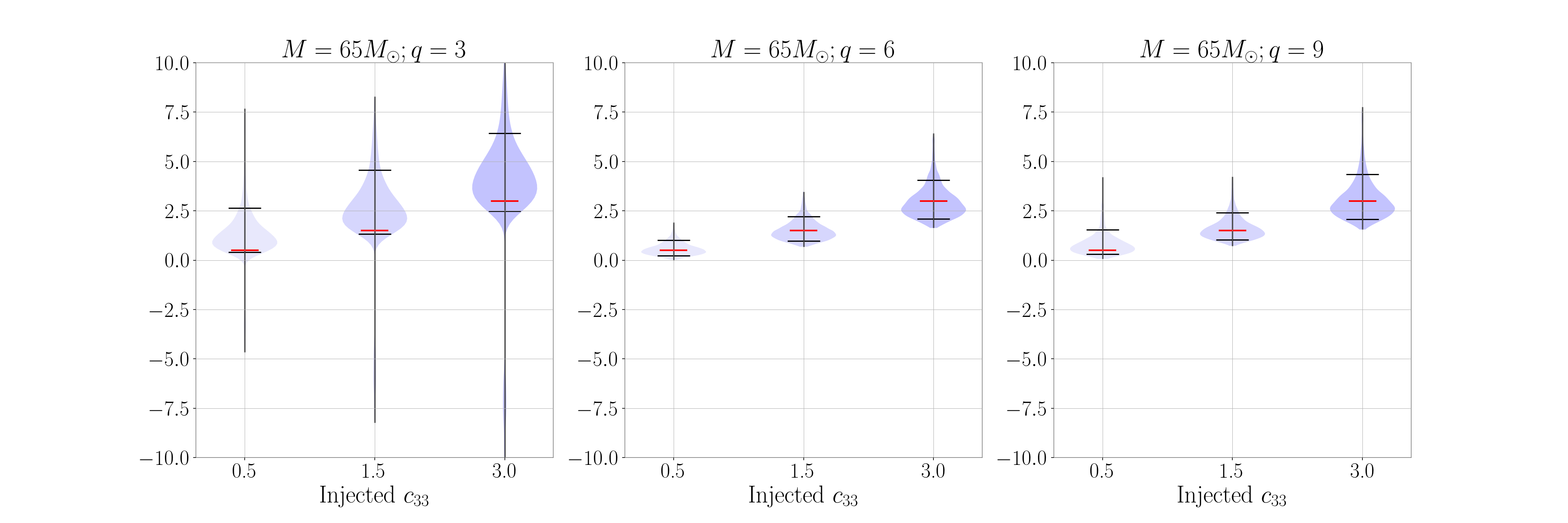}\hfill
\includegraphics[width=0.98\linewidth]{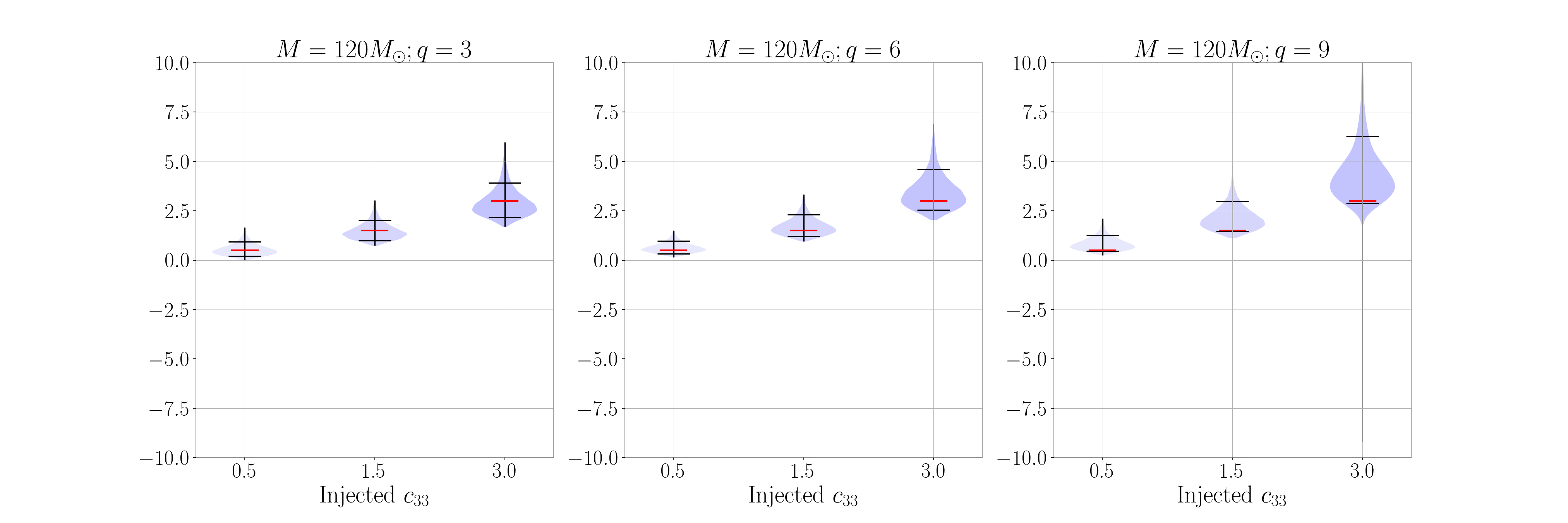}\hfill
\includegraphics[width=0.98\linewidth]{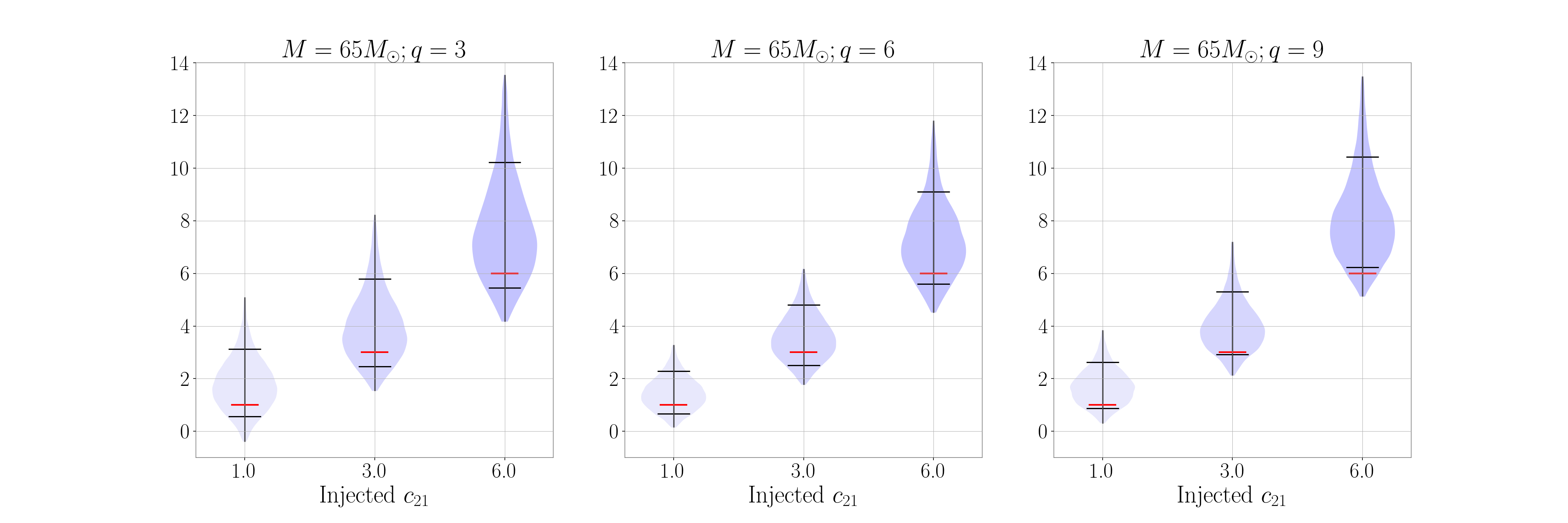}\hfill
\includegraphics[width=0.98\linewidth]{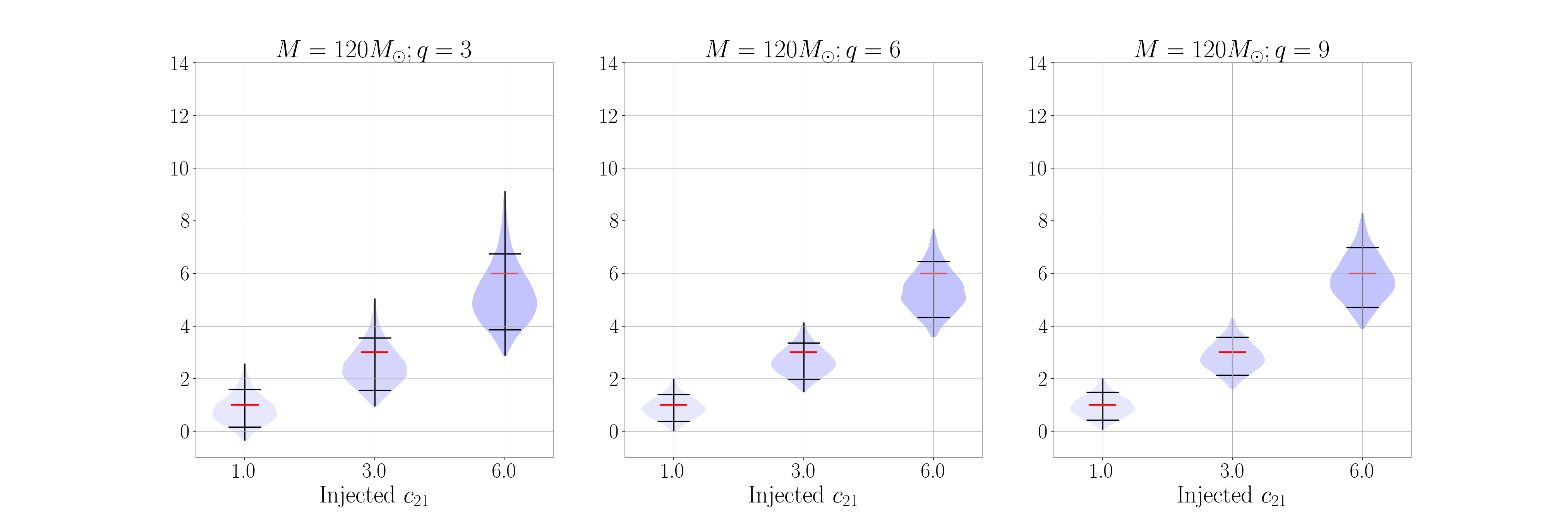}\hfill
 \caption{Violin plots for the posterior density distributions of $c_{33}$ (top two rows) and $c_{21}$
 (bottom two rows), for $M = 65, 120\,M_\odot$, and $q = 3$ (left column), 
 $q= 6$ (middle column), and $q  = 9$ (right column). In each case the black horizontal bars indicate
90\% confidence intervals, and the red horizontal bar the injected value; the black vertical 
line shows the support of the posterior.}
 \label{fig:m65_120_damp_recovery}
\end{figure*}

Fig.~\ref{fig:m65_120_damp_recovery} shows posterior probability densities 
for the corresponding injections. The trends show broad consistency with what we 
saw for the log Bayes factors. In particular, for the injected values $c_{33} = 0.5$
and $c_{21} = 1$, posterior densities either include the GR value of zero, or extend
to quite close to it, while for higher injected values, the GR value tends to be 
outside the support of the distribution. Also, the 90\% confidence intervals tend to 
be tighter for higher total mass and for higher mass ratio, again consistent with 
the behavior of the $\ln\mathcal{B}^{\rm NonGR}_{\rm GR}$ in Fig.~\ref{fig:bayes_factor_difference}, 
and indeed with Fig.~\ref{fig:various_modes_snr}.

\subsection{Injections with parameters similar to those of GW190412 and GW190814}

\begin{figure*}
\centering
\includegraphics[width=0.49\linewidth]{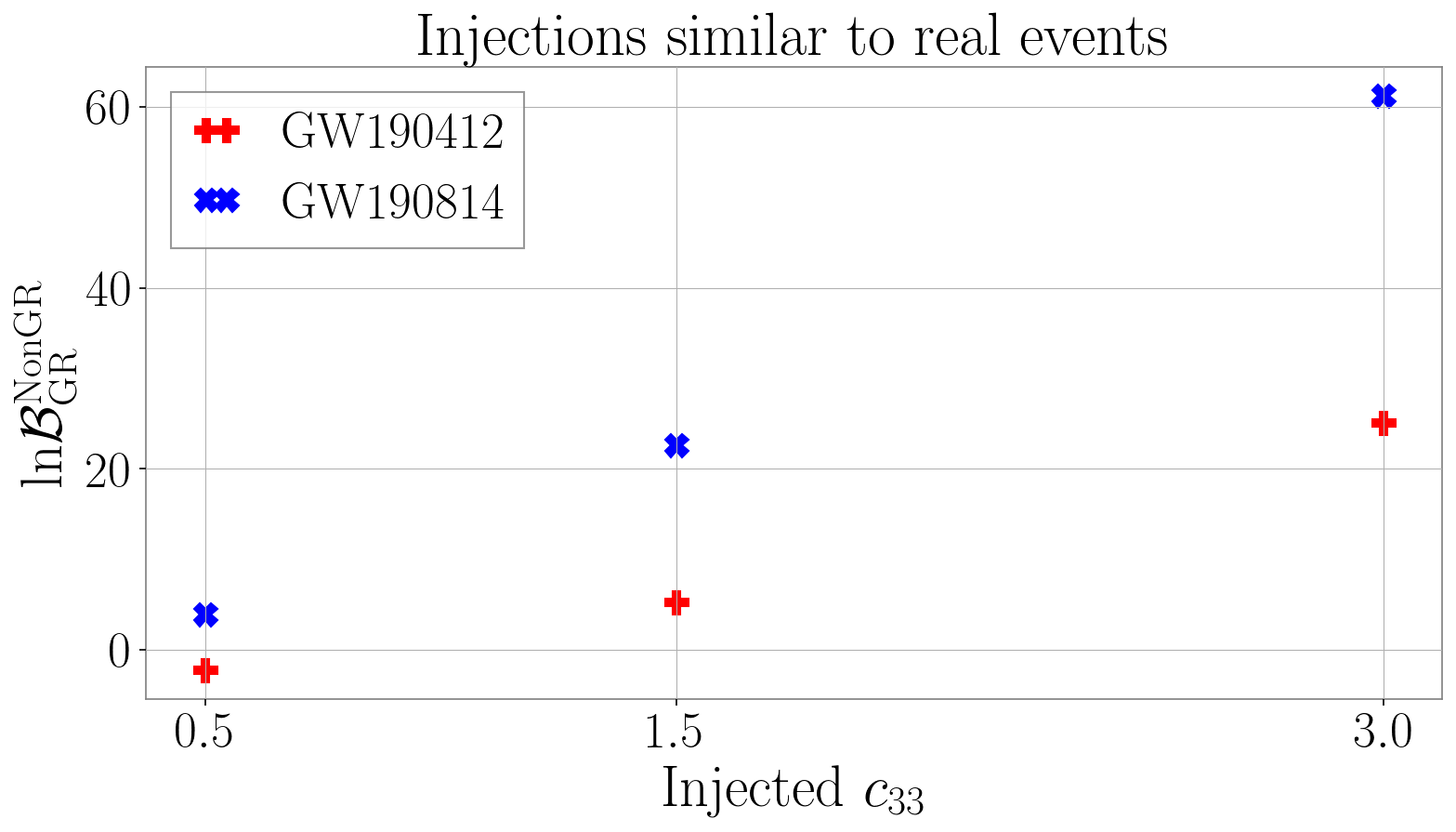}\hfill
\includegraphics[width=0.49\linewidth]{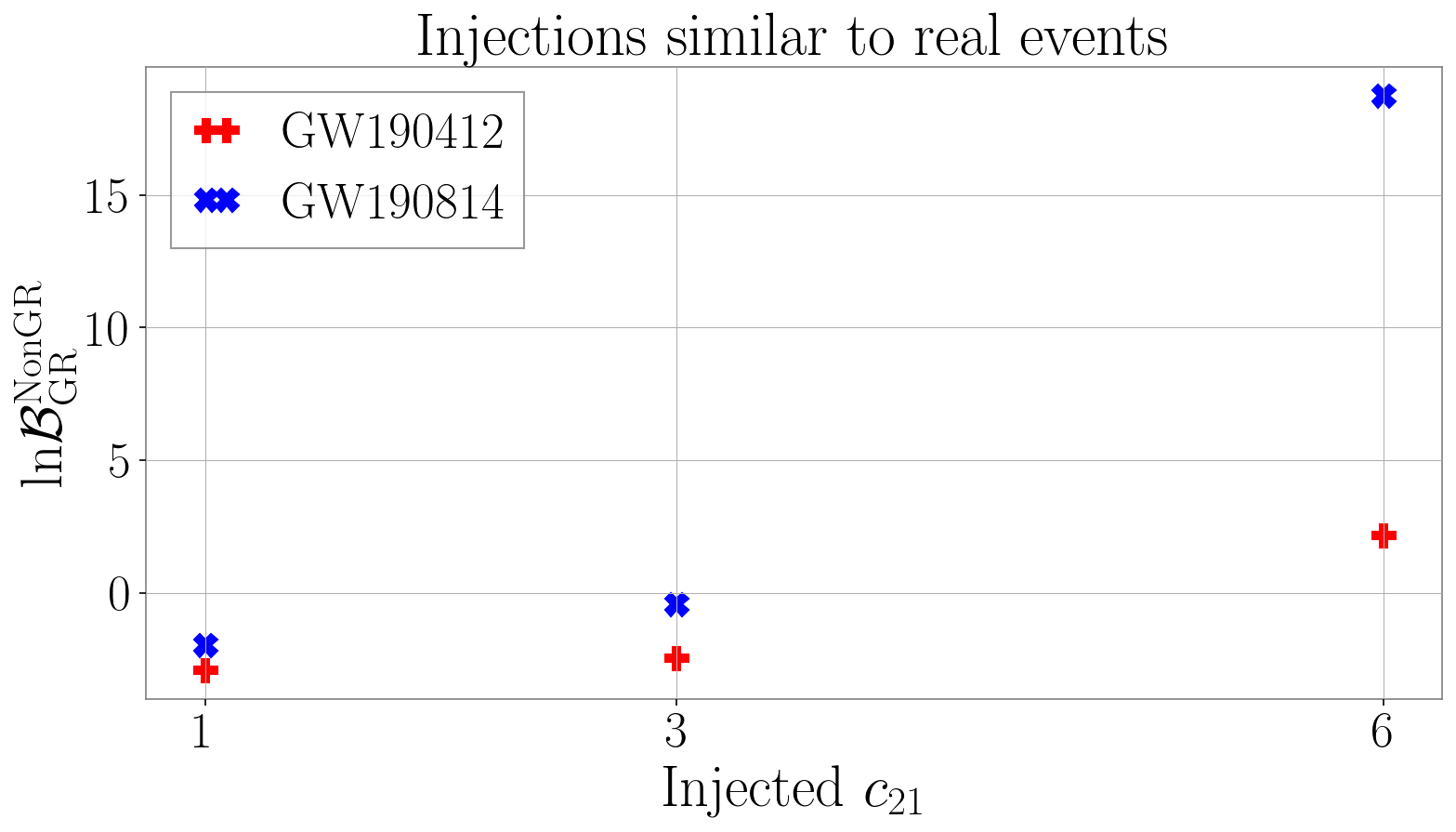}\hfill
\caption{$\ln\mathcal{B}^{\rm NonGR}_{\rm GR}$ for injections with GR parameters 
similar to those of GW190412 and GW190814.}
\label{fig:bayes_factor_difference_real_event}
\end{figure*}

Next we turn to injections with GR parameters close to those of the real events GW190412 and
GW190814. Fig.~\ref{fig:bayes_factor_difference_real_event} shows results for 
$\ln\mathcal{B}^{\rm NonGR}_{\rm GR}$. Here too the trends are as expected: the log Bayes factor
increases with increasing injected values for $c_{33}$ and $c_{21}$. Note that although 
GW190412 had a higher mass than GW190814 ($M = 46.6\,M_\odot$ versus $M = 27.6\,M_\odot$),
the mass ratio of GW190412 was considerably smaller than that of GW190814 ($q = 4.2$ versus 
$q = 9.3$). The log Bayes factors are higher
for the latter event, consistent with Fig.~\ref{fig:various_modes_snr}. 
We see that for GW190412 one has $\ln\mathcal{B}^{\rm NonGR}_{\rm GR} < 0$ for 
$c_{33} = 0.5$, and the same is true for both injections in the cases 
$c_{21} = 1$ and $c_{21} = 3$, presumably due to the lower total masses. 

\begin{figure*}
\centering
\includegraphics[width=0.99\linewidth]{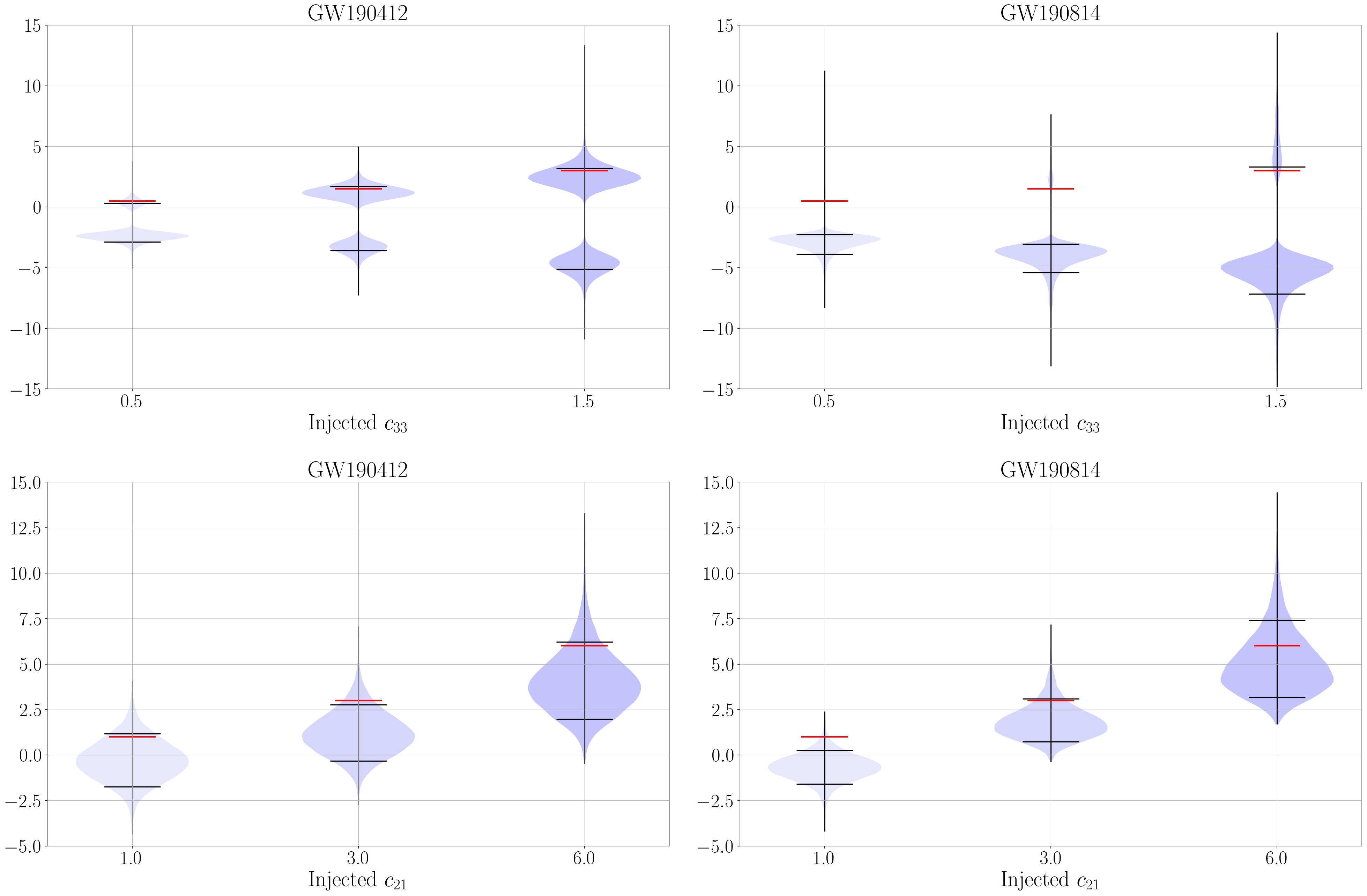}\hfill
\caption{Violin plots for the posterior density distributions of 
$c_{33}$ (top row) and $c_{21}$ (bottom row), for injections similar to  
GW190412 (left column) and GW190814 (right column). In each case the black horizontal bars indicate
90\% confidence intervals, and the red horizontal bar the injected value; the black vertical 
line shows the support of the posterior.}
\label{fig:posteriors_real_event}
\end{figure*}


Fig.~\ref{fig:posteriors_real_event} shows posterior probability distributions for the
same injections. In all cases, the injected value for $c_{33}$ and $c_{21}$ lies within 
the support of the posterior. For $c_{21}$ the results look like what one might expect, but 
for $c_{33}$ the posteriors are bimodal, with the true value not always lying in the strongest 
mode. As will be clarified in the next section, this behavior results from a partial degeneracy 
between $c_{33}$ and the inclination angle $\iota$.




\subsection{Results for GW190412 and GW190814}

Finally we turn to the real events GW190412 and GW190814. Table \ref{tab:logB} shows the 
results for $\ln\mathcal{B}^{\rm NonGR}_{\rm GR}$ when comparing the hypothesis of 
a non-zero $c_{33}$ or $c_{22}$ with the GR hypothesis. All the log Bayes factors are negative, 
so that we have no reason to suspect a violation of GR in the amplitudes of sub-dominant modes.

\begin{figure*}[h]
    \centering
    \includegraphics[width=16cm]{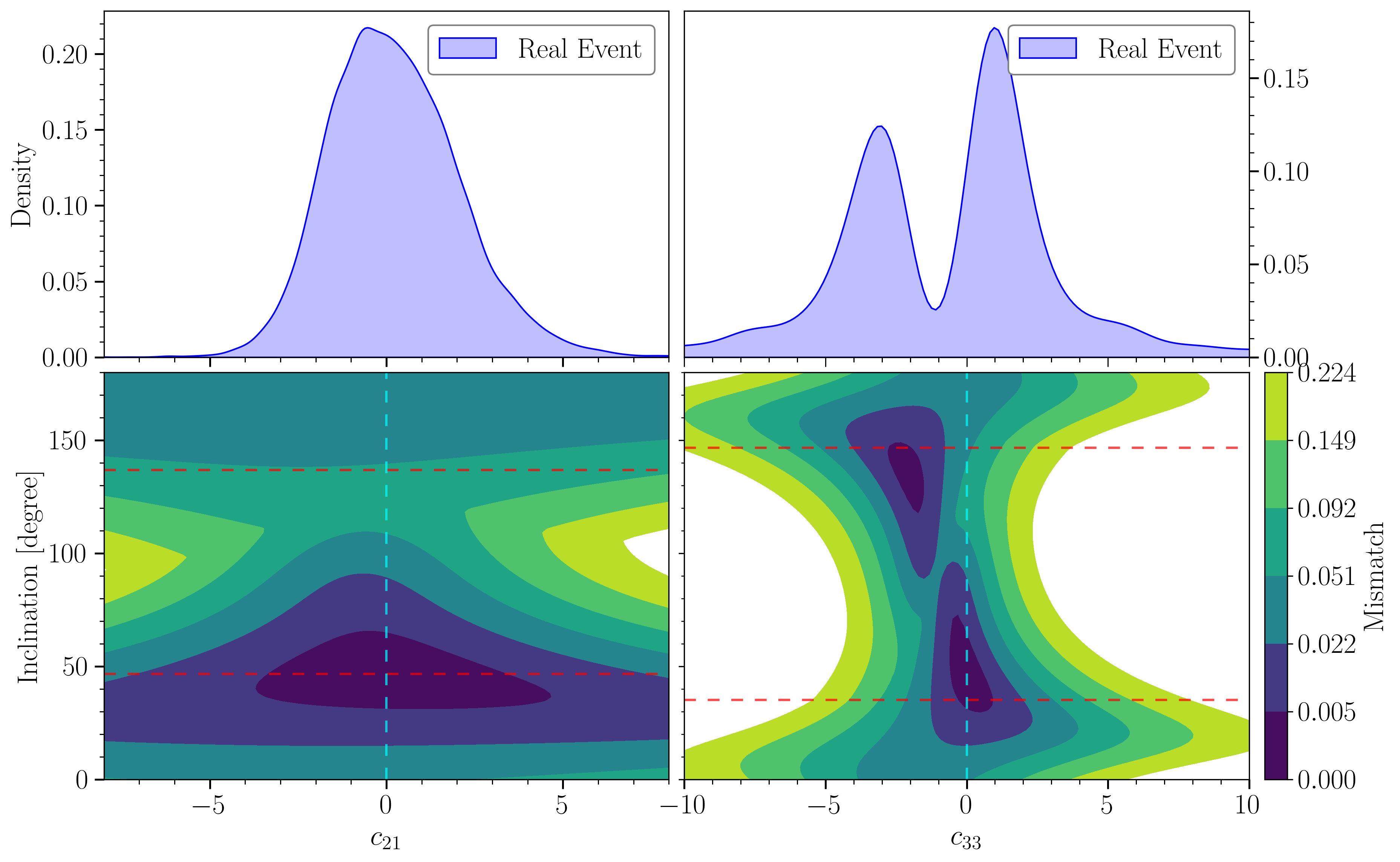}
    \caption{Top panels: Posterior density functions for $c_{21}$ (left) and $c_{33}$ (right)
    for GW190412. 
    Bottom panels: Contours of constant mismatch between the maximum-likelihood GR waveform, and 
    a waveform in which $\iota$ and $c_{21}$ (left) or $c_{33}$ (right) are varied while keeping all other 
    parameters the same. The dashed vertical lines indicate the GR values $c_{21} = 0$ and 
    $c_{33} = 0$, respectively, and the dashed horizontal lines indicate the peak-likelihood values for
    $\iota$ obtained from the analyses of GW190412 with respectively 
    $c_{21}$ and $c_{33}$ as free parameters.}
    \label{fig:posteriors_190412}
\end{figure*}

\begin{figure*}[h]
    \centering
    \includegraphics[width=16cm]{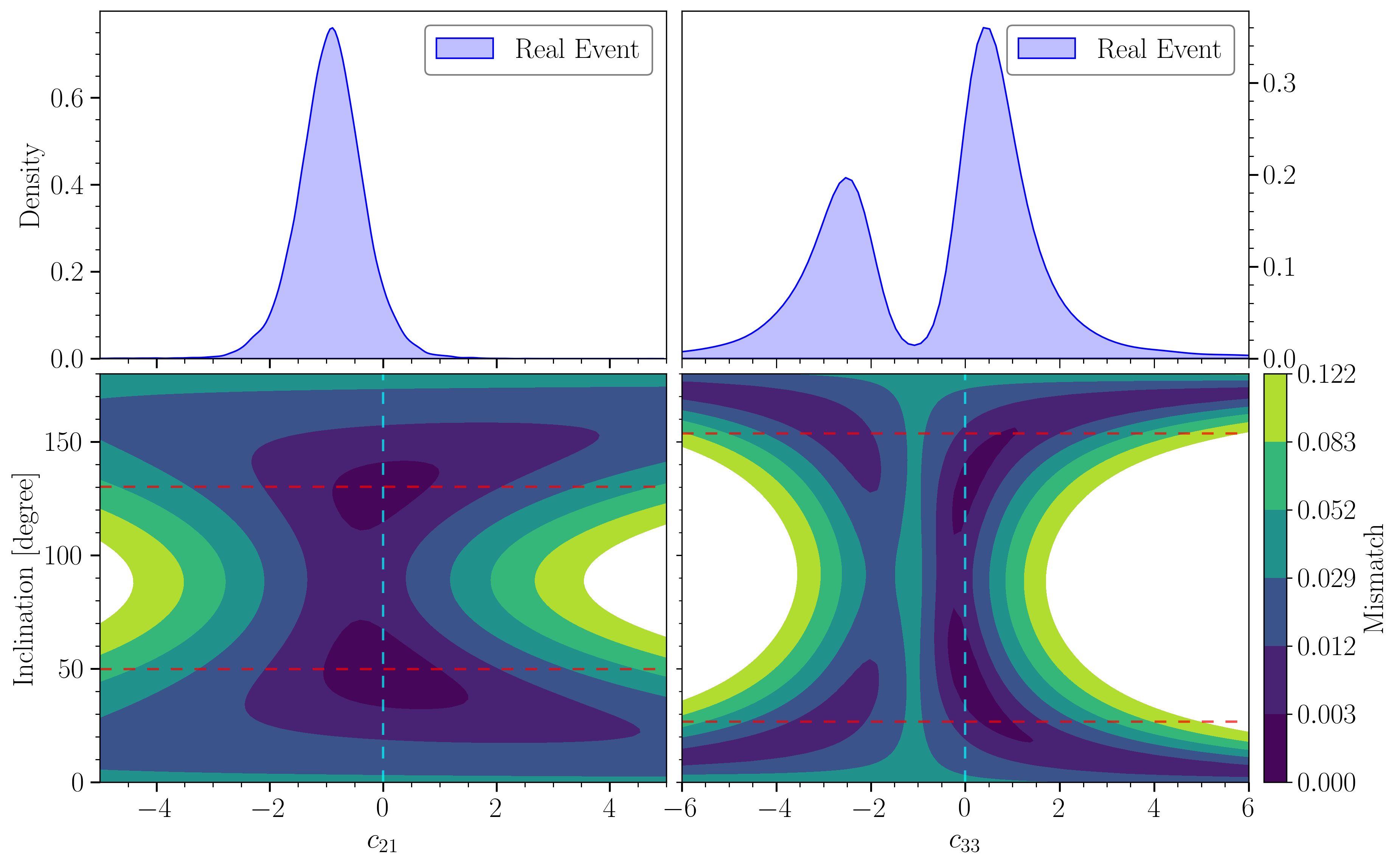}
    \caption{The same as in Fig.~\ref{fig:posteriors_190412} but for GW190814.}
    \label{fig:posteriors_190814}
\end{figure*}

\begin{table}
\begin{tabular}{ |c|c|c| }
\hline
Event & GW190412 & GW190814 \\
\hline
\hline
$c_{33}$ & -1.25 & -3.96 \\
\hline
$c_{21}$ & -2.48 & -1.77 \\ 
\hline
\end{tabular}
\caption{Values of $\ln\mathcal{B}^{\rm NonGR}_{\rm GR}$ for analyses of the real 
events GW190412 and GW190814.}
\label{tab:logB}
\end{table}


More interesting are the posterior distributions for $c_{21}$ and especially $c_{33}$, which are 
shown in Figs.~\ref{fig:posteriors_190412} and \ref{fig:posteriors_190814}. For both events, 
the posterior for $c_{21}$ is unimodal, and consistent with the GR value of zero. However, just 
like in the simulations of the previous section, the posterior for $c_{33}$ is bimodal, also 
for both events. 

As it turns out, this bimodality results from a degeneracy between $c_{33}$ and the inclination 
angle $\iota$. The lower panels of Figs.~\ref{fig:posteriors_190412} and \ref{fig:posteriors_190814}
show \emph{mismatches} between (a) a reference waveform $\tilde{h}_{\rm ref}(f)$, which is  
a GR waveform with maximum-likelihood parameters for the respective signals, and (b) a waveform
$\tilde{h}(c_{\ell m}, \iota; f)$ in which $c_{\ell m}$ and $\iota$ can take on arbitrary values, but 
all other parameters are the maximum-likelihood ones from the GR analysis. Specifically, 
we compute
\begin{equation}
\mbox{MM} = 1 
- \max_{t_{\rm ref}, \varphi_{\rm ref}}\frac{\langle h_{\rm ref} | h(c_{\ell m}, \iota)\rangle}{\sqrt{\langle h_{\rm ref}|h_{\rm ref}\rangle}\sqrt{\langle h(c_{\ell m}, \iota) |h(c_{\ell m}, \iota) \rangle}},
\end{equation}
where the maximization is over the above mentioned reference time and phase of the waveform, 
and we focus on $(\ell, m) = (2,1)$ and $(\ell,m) = (3,3)$. 

In the bottom panels of 
Figs.~\ref{fig:posteriors_190412}, \ref{fig:posteriors_190814}, these mismatches are indicated 
with color coding, with dark colors signifying small mismatch. Overlaid are dashed lines indicating 
the peak-likelihood values in the (bimodal) posterior distribution for $\iota$ obtained when analyzing 
the events with either $c_{21}$ or $c_{33}$ as additional free parameters. Focusing first on the case of 
$c_{33}$ and GW190412 
in Fig.~\ref{fig:posteriors_190412}, we see that there are two regions in the $(c_{33}, \iota)$
plane where mismatches are low: one region that contains the GR value $c_{33} = 0$ and is consistent
with the lower value of $\iota$, and another region consistent with the higher $\iota$ value 
and $c_{33} \neq 0$. In either region, waveforms $h(c_{33}, \iota)$ are consistent with the 
reference waveform $h_{\rm ref}$, which explains the bimodality in the posterior for $c_{33}$. 
By contrast, based on the analogous plot for $(c_{21}, \iota)$, no such bimodality is to be expected, 
and indeed, the posterior for $c_{21}$ is unimodal. The corresponding Fig.~\ref{fig:posteriors_190814}
for GW190814 leads to similar conclusions.

\section{Summary and conclusions}
\label{sec:conclusions}

We have set up a Bayesian analysis framework to test GR by looking at the amplitudes of
sub-dominant modes in GW signals from BBH coalescences, using a state-of-the-art waveform model.
Specifically, we allow for modifications in the amplitudes of the $(3,3)$ and $(2,1)$ modes, 
which tend to be the strongest among the sub-dominant modes. Apart from performing 
parameter estimation on the associated testing parameters $c_{33}$ and $c_{21}$, this allows 
for hypothesis ranking between the presence and absence of such anomalies in the modes. 

Results from simulations involving injected waveforms in stationary, Gaussian noise largely follow 
the trends one would expect based on the dependence of mode amplitudes on total mass and mass ratio: 
for similar SNRs, heavier and more asymmetric systems make it easier to find violations of GR 
of the type studied here. 

We then performed the first analysis of this kind on the real events GW190412 and GW190814, 
which were associated with significantly unequal component masses, and in which strong evidence 
for sub-dominant mode content had been found 
\cite{LIGOScientific:2020stg,LIGOScientific:2020zkf,Roy:2019phx}. Log Bayes factors 
indicated no evidence for a GR violation in either the $(2,1)$ or $(3,3)$ mode. In the case 
where the $(3,3)$ mode was being investigated, the posterior density function for 
$c_{33}$, while being consistent with the GR value $c_{33} = 0$, did exhibit bimodality, 
but this was shown to result from correlations between $c_{33}$ and the inclination angle 
$\iota$. Since the bimodality was also present in $c_{33}$ posterior densities for injections with 
parameters similar to the ones of GW190412 and GW190814 and $c_{33} \neq 0$, some caution is
called for in interpreting such posteriors, at least for BBHs with total mass $M \lesssim 50\,M_\odot$. 
However, our results show that log Bayes factors $\ln\mathcal{B}^{\rm NonGR}_{\rm GR}$, 
which were not considered in previous work in this context \cite{Islam:2019dmk}, are robust 
indicators for or against the presence of a violation of GR. 

Even in systems with significantly asymmetric masses and high total mass, with second-generation 
detectors, GR violations have to be sizeable ($c_{33} \gtrsim 1.5$ and $c_{21} \gtrsim 3$) 
in order to be confidently detected. It will be of interest to see how the sensitivity 
of our method will improve going towards Einstein Telescope 
\cite{Punturo:2010zza,Hild:2010id,Maggiore:2019uih,Kalogera:2021bya}, Cosmic 
Explorer \cite{LIGOScientific:2016wof,Reitze:2019iox,Evans:2021gyd,Kalogera:2021bya}, 
and the space-based LISA \cite{Babak:2021mhe}, but this is left for future work.

\begin{acknowledgments} 

  A.P., C.K., S.R., Y.S., and C.V.D.B.~are supported by the research programme 
  of the Netherlands Organisation for Scientific Research (NWO). I.G. and B.S.S.~are 
  supported by NSF grant numbers PHY-2012083 and AST-2006384.
  The authors are grateful for computational resources provided by the 
  LIGO Laboratory and supported by the National Science Foundation Grants No.~PHY-0757058 and 
  No.~PHY-0823459. 
  This
  research has made use of data, software and/or web tools
  obtained from the Gravitational Wave Open Science Center (https://www.gw-openscience.org), a service of LIGO
  Laboratory, the LIGO Scientific Collaboration and the
  Virgo Collaboration. LIGO is funded by the U.S. National Science Foundation. Virgo is funded by the French
  Centre National de Recherche Scientifique (CNRS), the
  Italian Istituto Nazionale della Fisica Nucleare (INFN)
  and the Dutch Nikhef, with contributions by Polish and
  Hungarian institutes.
  
\end{acknowledgments}

\bibliography{reference}{}

\end{document}